\documentclass[twocolumn,showpacs,showkeys,
preprintnumbers,
amsmath,amssymb,page]{revtex4}
\usepackage{graphicx}
\usepackage{dcolumn}
\usepackage{bm}

\begin{document}


\title{Irregular Dynamics in a Solvable One-Dimensional Quantum Graph}

\author{
Pavel Hej{\v c}{\'i}k
  and
Taksu Cheon
}
\affiliation{
Laboratory of Physics,
Kochi University of Technology,
Tosa Yamada, Kochi 782-8502, Japan
}

\date{February 27, 2005}

\begin{abstract}
We show that the quantum single particle motion on a one-dimensional line 
with F{\"u}l{\"o}p-Tsutsui point interactions exhibits characteristics 
usually associated with nonintegrable systems 
both in bound state level statistics and scattering amplitudes.
We argue that this is a reflection of  the underlying stochastic dynamics which 
persists in classical domain.
\end{abstract}

\maketitle

%
The advancement in nano-engineering in the last decade 
has brought novel incentives to the study of low-dimensional 
quantum systems with geometrically designed forms that have
no counterpart in nature. 
The quantum graph, which is a generic one-dimensional model of nano-device 
composed of quantum wires, represents one of such systems \cite{EX96,KS99}.
The interest to the quantum graph is enhanced with its possibility to
emulate the two-dimensional system of
quantum billiard \cite{EH05},
whose solution has required rather extensive numerical treatments.
It is therefore quite appropriate, at this point, to investigate generic aspects and
general features of quantum graphs ahead of detailed studies of specific 
models of nano-devices.

In a parallel development, quantum graphs have been used
as a tractable model for the study of quantum chaos, or
the irregular aspects of quantum dynamics occurring as quantum manifestations of
classically chaotic systems \cite{KS97,KS00}.  
Naturally, it is expected that
random quantum graphs, which are complex networks of quantum lines,
would result in the universal quantum fluctuation
that has been associated to the quantum chaotic dynamics \cite{KS01,GA04}.
It has been revealed, however, that no real complex network is required
for the irregular quantum dynamics to present itself.
A very simple version of the quantum graph,
the star graph, which is a quantum graph with  many lines connected at a single node,
has been shown to display the characteristics associated to partial quantum chaos
in an analytical semiclassical study \cite{BK99}.
%
A natural question to be asked is whether we can further simplify
the model of quantum chaos to the point of solvability.
In this article, we consider one of the simplest possible quantum graph
which is made up
solely of nodes with two connected lines with the special property for 
the nodes called scale invariance.  
The resulting system amounts to a single one-dimensional line 
with number of scale invariant point interactions.
We show that the system has elementary analytical scattering matrices and
also elementary analytical eigenvalue equation, yet displays
full characteristics of irregular quantum dynamics, both in scattering
amplitudes and in  bound state level statistics.
We discuss the implication of the results,
and look into
the apparent contradiction 
of the appearance of the quantum chaos in
a  seemingly integrable,
solvable conservative one-dimensional system.
%

%
\begin{figure}
\center{
\includegraphics[width=8cm]{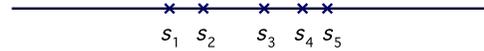}
}
\label{fig1}
\caption
{
A schematic depiction of our model system.  
A free quantum particle moves along the line on which
point interactions described by
(3), (4) and (6) are located at $s_1, \cdots, s_N$.  This example
shows the case of $N=5$. 
}
\end{figure}
%
%
We consider a quantum particle, constrained to move on a one-dimensional line
with $N$ point-like defects \cite{AG88}  whose locations are given by
$x=s_i$ with $i = 1, 2, .. N$ (FIG. 1).
The Hamiltonian of the system is given, in appropriately rescaled unit, by
\begin{eqnarray}
& &H = -\frac{1}{2} \frac{d^2}{d x^2}
\quad{\rm on}
\\ \nonumber
& &
x \in (-\infty, s_1) \cup (s_1, s_2)  \cup ...  (s_{N-1}, s_N) \cup (s_N, \infty).
\end{eqnarray}
The dynamics of the system is described by the Schr{\" o}dinger equation
\begin{eqnarray}
H \psi(x) = E \psi(x),
\end{eqnarray}
supplemented by the $U(2)$ connection conditions 
at the defects \cite{SE86}, which is conveniently specified \cite{FT00} by 
\begin{eqnarray}
\label{FTcnd}
(U_i-I) \Phi(s_i) + i L_0 (U_i+I) \Phi^\prime(s_i) = 0,
\end{eqnarray}
where $i$ runs as $i=1, 2, ..., N$, and
$U_i$ is a unitary matrix belonging to $U(2)$ group. 
The boundary vectors $ \Phi(x)$ and $ \Phi^\prime(x)$ are given by
\begin{eqnarray}
\Phi(s_i) = \begin{pmatrix} \psi_+(s_i) \\ \psi_-(s_i) \\  \end{pmatrix},
\quad
\Phi^\prime(s_i) = 
\begin{pmatrix} \psi^\prime_+(s_i)\cr -\psi^\prime_-(s_i) \end{pmatrix}
,
\end{eqnarray}
where  $\psi_\pm(s_i)$ and $\psi^\prime_\pm(s_i)$  denote
the limit value of $\psi(x)$ and its derivative
from the upper and lower regions of  the defects $s_i$,
$x \to s_i\pm0$.  The constant $L_0$ is a length scale introduced
to account for the scale anomaly \cite{JA95}
inherent in one-dimensional point
interaction.
%
%
For technical simplicity, we assume all $U_i$ to be identical, $U = U_i$.
Among all possible $U$, we consider {\it scale invariant} subfamily, discovered by
F{\"u}l{\"o}p and Tsutsui \cite{FT00}
whose $U$ has property
\begin{eqnarray}
{\rm det}[U \pm I] = 0 .
\end{eqnarray}
This condition guarantees the equation (\ref{FTcnd}) without any involvement 
of the scale
parameter $L_0$.
The standard parametrization for this class of $U$ is
\begin{eqnarray}
\label{USI}
U 
= \begin{pmatrix}\cos \theta & e^{i\phi}\sin\theta \cr
   e^{i\phi}\sin\theta  &  -\cos \theta \end{pmatrix} .
\end{eqnarray}
This gives  the connection condition which reads
\begin{eqnarray}
\label{BCSI}
& &\frac{e^{i\phi}}{\alpha}\psi_-(s_i) = \psi_+(s_i),
\nonumber\\
& &e^{i\phi}\alpha\psi^\prime_-(s_i) = \psi^\prime_+(s_i),
\end{eqnarray}
where the ``strength'' $\alpha$ is defined by 
\begin{eqnarray}
\alpha = - \cot \frac{\theta}{2}.
\end{eqnarray}
The F{\"u}l{\"o}p-Tsutsui point interaction (\ref{USI}) is a less known
subclass of one-dimensional point interaction
compared to the standard $\delta$ potential and $\delta^\prime$ 
(or $\varepsilon$) potential, but its property of scale invariance comes
in handy in our following treatment.  
We stress that this seemingly exotic interaction is nevertheless realizable
as a short-range limit of certain local potential \cite{CS98}.
%
%
We first consider the scattering by a single defect.    
Let us, for a moment,  suppose that there is only a single defect located at $x=s$.
Considering the incoming wave from $x < s_1$ side, 
we assume the wave function to be in the form 
\begin{eqnarray}
\psi(x) 
&=& e^{i k x} - R_1(s_i)  e^{-i k x} \quad (x<s_i) ,
\nonumber \\
&=& T_1(s_i)  e^{i k x} \qquad\qquad\ \  (x>s_i) .
\end{eqnarray}
We obtain the transmission and reflection amplitudes as
\begin{eqnarray}
T_1(s_i) = \frac{2\alpha}{1+\alpha^2} e^{i\phi} ,
\quad
R_1(s_i) = \frac{1-\alpha^2}{1+\alpha^2} e^{2i k s_i}.
\end{eqnarray}
For the scattering from $x>s_N$ side, we write
\begin{eqnarray}
\psi(x) 
&=& T^\prime_1(s_i)  e^{-i k x} \qquad\qquad  (x>s_i) .
\nonumber \\
&=& e^{-i k x} - R^\prime_1(s_i)  e^{i k x} \quad (x<s_i) ,
\end{eqnarray}
and obtain
\begin{eqnarray}
T^\prime_1(s_i) = T^*_1(s_i),
\quad
R^\prime_1(s_i) = -R^*_1(s_i).
\end{eqnarray}
The absence of the scale parameter $L_0$ results in the energy independence of 
scattering amplitude.

With elementary algebra, we can write the scattering amplitudes for $N$ defects 
in the recursive forms; For the left-right amplitudes $T_N$ and $R_N$, we have
\begin{eqnarray}
\label{SCG}
& &T_N(s_1,...,s_N) 
=  \frac{T_1(s_1)T_{N-1}(s_2,...,s_N)}{1+R_1^*(s_1)R_{N-1}(s_2,...,s_N)} ,
\qquad
%
\nonumber \\ 
\\ \nonumber
%
& &
R_N(s_1,...,s_N) 
=  \frac{R_1(s_1)+R_{N-1}(s_2,...,s_N)}{1+R_1^*(s_1)R_{N-1}(s_2,...,s_N)} .
\\ \nonumber
\end{eqnarray}
The right-left amplitude $T^\prime$ $R^\prime$ are obtained from
\begin{eqnarray}
& &T^\prime_N(s_1, ..., s_N) = T_N^*(s_N, ..., s_1) ,
\nonumber \\
& &R^\prime_N(s_1,  ..., s_N) = -R_N^*(s_N, ..., s_1) .
\end{eqnarray}
Note the reversed ordering of $s_i$ in right and left hand sides of the equations.
With repeated iteration, we obtain explicit expressions for
scattering matricesin the form
\begin{eqnarray}
\label{SCE}
T_N(k) = \frac{\gamma^N  }{D_N(k)} ,
\quad
R_N(k)  = \frac{B_N(k)} {D_N(k)} ,
\end{eqnarray}
where $B_N(k)$ and $D_N(k)$ are defined by
\begin{widetext}
\begin{eqnarray}
\label{SCECD}
& &\!\!\!\!\!\!\!\!
B_N(k) =
{\beta  \sum_{i}^{N} e^{2ik s_i} 
    +\beta^3 \!\!\!\! \sum_{i>j>m}^{N} \!\!\!\! e^{2ik (s_i-s_j+s_m)} 
    +\beta^5 \!\!\!\!\!\!\!\!\!\! \sum_{i>j>m>n>p}^{N} \!\!\!\!\!\!\!\!\!\!
     e^{2ik (s_i-s_j+s_m-s_n+s_p)}  +\cdots
} ,
\\
& &\!\!\!\!\!\!\!\!
D_N(k) =
{1
    +\beta^2  \sum_{i>j}^{N} e^{2ik (s_i-s_j)} 
    +\beta^4 \!\!\!\!\!\! \sum_{i>j>m>n}^{N} \!\!\!\!\!\! e^{2ik (s_i-s_j+s_m-s_n)} 
    +\beta^6 \!\!\!\!\!\!\!\!\!\!\!\! \sum_{i>j>m>n>p>q}^{N} \!\!\!\!\!\!\!\!\!\!\!\!
     e^{2ik (s_i-s_j+s_m-s_n+s_p-s_q)}  +\cdots
} ,
\end{eqnarray}
\end{widetext}
%
and the abreviations 
\begin{eqnarray}
\beta = \frac{1-\alpha^2}{1+\alpha^2}, \quad 
\gamma = \frac{2\alpha}{1+\alpha^2} e^{i \phi}
\end{eqnarray}
are used.  The sum runs over all indices in the range between $1$ and $N$ with
the specified constraint, and the numerator contains terms up to the order of
$\beta^{[N/2]}$ where the exponent signifies the integer part of $N/2$.
For given $N$, there are ${}_NC_l$ terms with order $\beta^l$, and 
the scattering matrices are the multi-periodic oscillatory functions 
with $2^{N-1}$ frequencies.
Along with scattering, we can also consider the bound spectra
by limiting the system to finite line of size $L \ge s_N$.
One of the easiest way is to impose Dirichlet boundary conditions at  $x=-L$ and $x=L$, 
$\psi(L)=\psi(-L) = 0$.
This leads, for the case of $\phi=0$, to the eigenvalue equation
%
\begin{eqnarray}
\label{BOG}
\left(R_N(k) - \! e^{-2i k L} \! \right)  \left( R^\prime_N(k) -\! e^{-2i k L}\! \right)
= T_N(k) T^\prime_N(k).
\end{eqnarray}
%
Explicite calculation again yields the form
\begin{widetext}
\begin{eqnarray}
\label{BOE}
& &\!\!\!\!
 \sin 2kL
+\beta \sum_{i}^{N}\sin  2k s_i
+\beta^2 \sum_{i>j}^{N}\sin  2k (s_j \!-\! s_i\!+\!L) 
+\beta^3 \!\!\!
\sum_{i>j>m}^{N}\!\!\!\!\! \sin  2k (s_m \!-\! s_j \!+\! s_i) 
\nonumber \\
& & \qquad
+\beta^4 \!\!\!\!\!
\sum_{i>j>m>n}^{N}\!\!\!\!\! \sin 2k (s_n \!-\! s_m \!+\! s_j \!-\! s_i\!+\!L) 
+\beta^5 \!\!\!\!\!
\sum_{i>j>m>n>p}^{N}\!\!\!\!\! \sin 2k (s_p \!-\! s_n \!+\! s_m \!-\! s_j+\! s_i) 
+\cdots  = 0,
\end{eqnarray}
\end{widetext}
which is a bound state counterpart of (\ref{SCE}).

%
In order to reveal the physical content of the scattering matrices (\ref{SCG})
and the spectral function (\ref{BOG}),
we plot $|T_N|^2$ as the function of incident momentum $k$ 
for  value of $\alpha=3/2$ in FIG.2. 
The number of point defects is
set to be $N=3$, $N=5$ and $N=7$. 
The angle $\phi$ is set to be zero for all cases.   
The locations $s_i$ are set to be the sum of square root of
primes; $s_i=\sum_{j=1}^i\sqrt{p_j}$,  where $p_j$ stands for the $j$-th
smallest prime with convention $p_1=1$, $p_2=2$.  
These values are selected to guarantee the incommensurability of $s_i$.
This also models a generic case of random sequencing of
successive $s_i$.  We have checked that different choice of $s_i$,
different ordering of relative size $s_{i+1}-s_{i}$ does not alter 
the essential characteristics of the results.
%
%
\begin{figure}
\center{
\includegraphics[width=8cm]{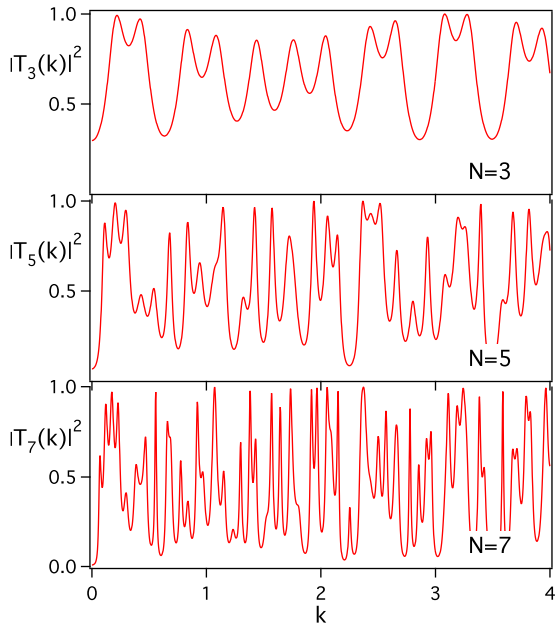}
}
\label{fig2}
\caption
{
Transmission probability as the function of incident momentum.
A common strengths parameter $\alpha=3/2$ is adopted. 
The location of the points are chosen to be $s_1=1$, $s_2=s_1+\sqrt{2}$,
$s_3=s_2+\sqrt{3}$, $s_4=s_3+\sqrt{5}$ and
$s_5=s_4+\sqrt{7}$, $s_6 = s_5 + \sqrt{11}$ and $s_7 = s_6 + \sqrt{13}$.}
\end{figure}

Despite the very simple analytic expression (\ref{SCE}) of the scattering amplitude,
as we increase $N$, the scattering quickly acquires 
``quantum chaotic'' features \cite{BS88}, which is characterized by Ericson fluctuation \cite{ER60},
or the wild oscillation in scattering amplitudes caused by the overlapping resonances.
Because of the scale invariance of the each point interactions, the fluctuation appears 
in arbitrary energy scale.
Clearly, this fluctuation is the result of interferences among
multi-periodic oscillations with incommensurate frequencies, whose
number of periods  $2^{N-1}$ proliferates very fast with increasing $N$.
%

%
%
\begin{figure}
\includegraphics[width=8.0cm]{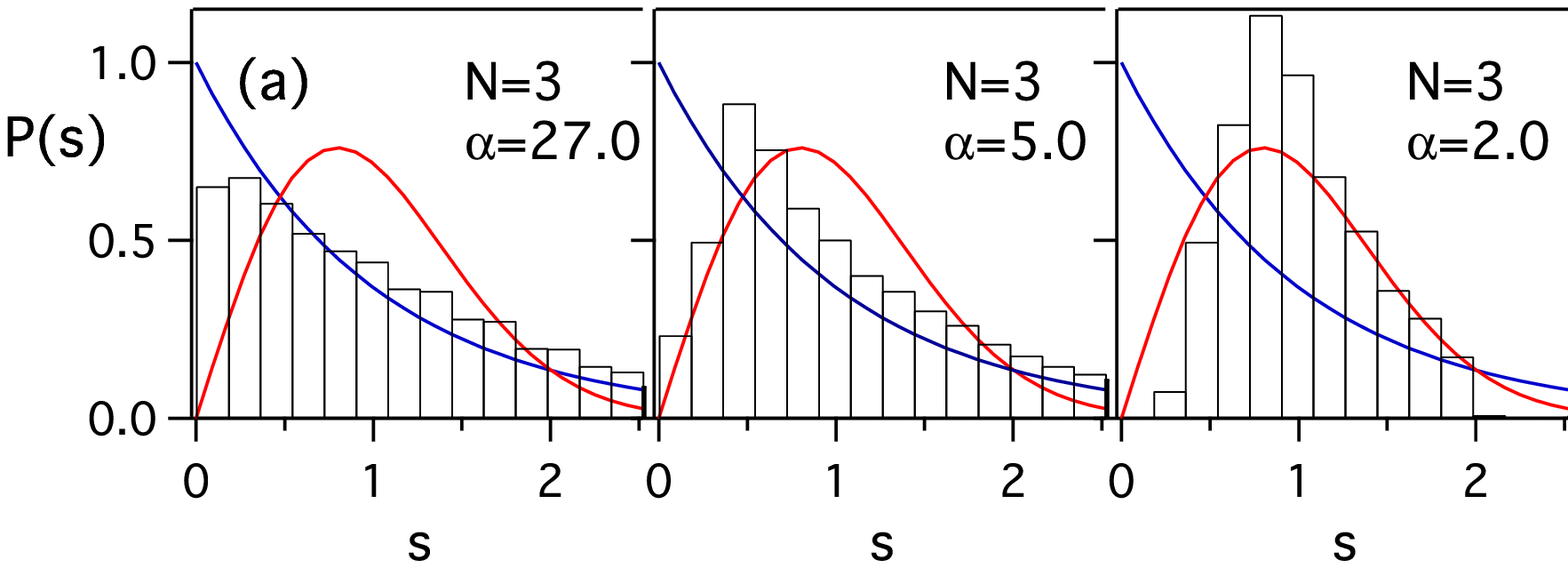}
\includegraphics[width=8.0cm]{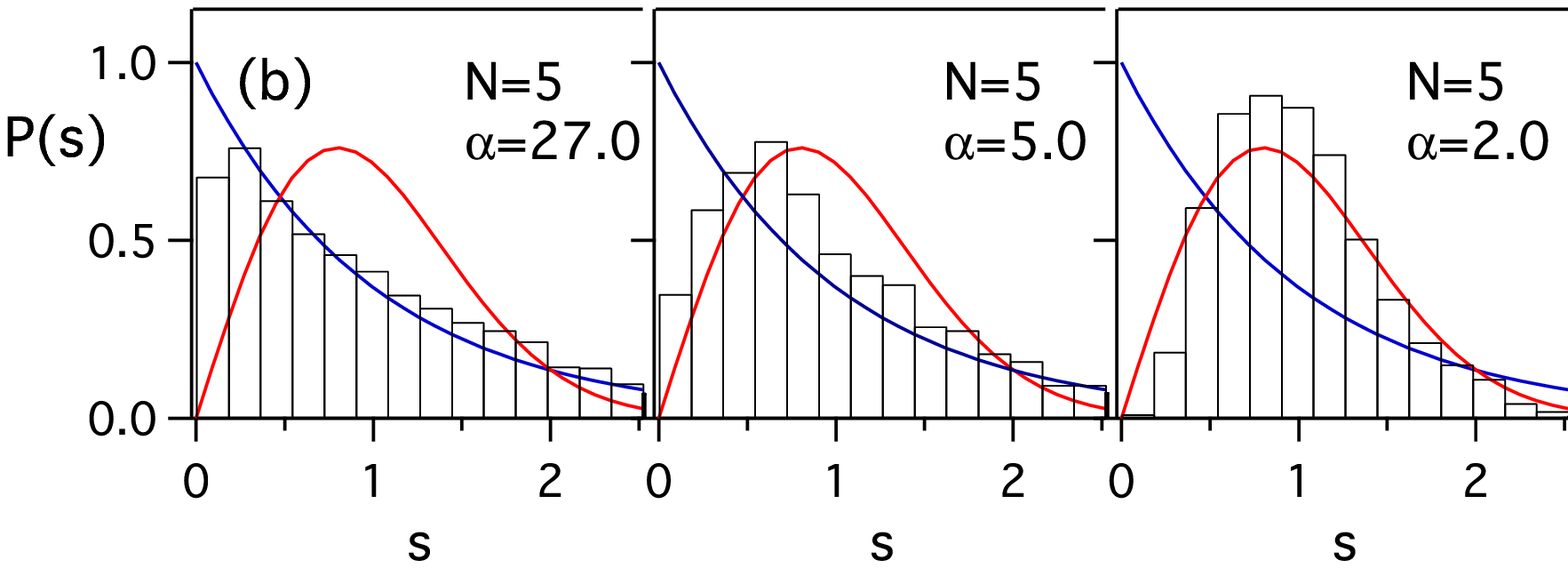}
\includegraphics[width=8.0cm]{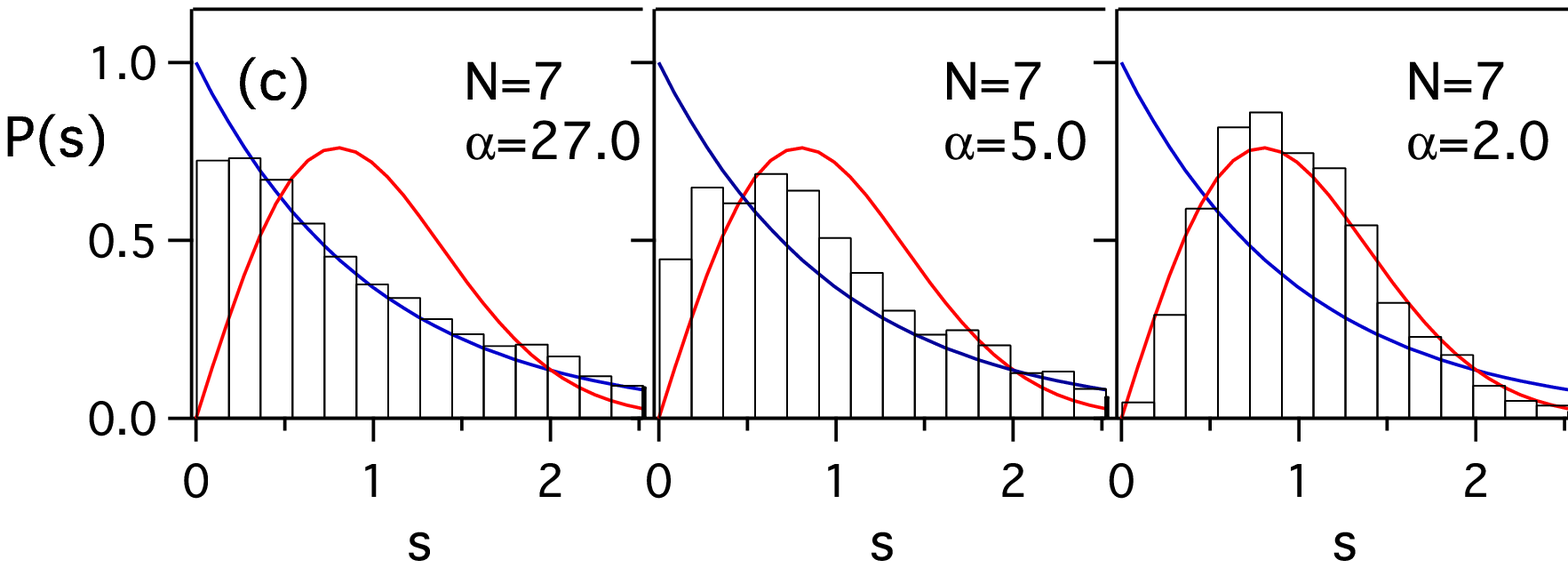}
\label{fig3}
\caption
{
Nearest neighbor spacing distribution $P(s)$ obtained from the
scaled energy levels of one-dimensional line limited to 
$x=[-L,L]$ with scale
invariant point interactions. Location parameters $s_i$ are same, 
apart from the re-scaling, as in FIG. 2.
Dirichlet boundary condition is imposed at both ends, $x=-L$ and $L$.
The solid line represent the Wigner distribution 
$P(s) = \pi/2 \cdot s \exp({-\pi/4 \cdot s^2})$,
and the dashed line
the Poisson distribution $P(s) = \exp({-s})$.
}
\end{figure}
%
%
We next examine the statistical properties of energy level sequence
calculated from our system with periodic boundary condition.
In FIG. 3, level spacing distribution $P(s)$ for the system with several $\alpha$
are shown for $N=3$, $N=5$, and $N=7$ cases from top to bottom 
as FIG3(a), FIG3(b) and FIG3(c). 
The distances $s_i$ are chosen to be
$s_i-s_{i-1} : s_{i+1}-s_i = \sqrt{p_i} : \sqrt{p_{i+1}}$.
The total length $L$ is set to be $s_1 : 2L$
$  = 1 : \sum_{i=1}^{N+1}p_i$.
We have chosen $\alpha=27$ to approximate disconnected large coupling limit,
and $\alpha=2$ as the strong coupling limit, while, as an intermediate
coupling, we chose the value $\alpha=5$. 
These graphs clearly show  the approach of $P(s)$
to the Wigner distribution
(also known as GOE distribution), which is regarded as the quantum signature 
of classically chaotic dynamics \cite{BG84},
at the strong coupling value as we increase the number of points $N$.
The convergence appears to be fast
as the Wigner-like level statistics already takes shape even with $N=3$.
We have confirmed, with numerical calculations up to $N=19$, that 
very good convergence to the Wigner statistics is obtained with large $N$ if the
coupling parameter $\alpha$ is properly rescaled to ensure the 
partial reflection/transmission of the system.

%
We now discuss the implication of our findings in a broader context.
The central result of this article is the generation of  the ``random'' or irregular
properties in quantum dynamics out of fully analytic 
quantum spectral functions obtained from a one-dimensional system. 
This type of properties are usually associated to
the  nonintegrable system.  
Since the classical counterpart of conservative one-dimensional system
necessarily is integrable, 
the well-established correspondence between the chaotic classical dynamics
and the random quantum dynamics seems to fail for our model.
The clue to understand this seeming contradiction might be found in
the singular nature of the high-energy limit of our system.
Because of the special property of scale invariance present 
in  F{\"u}l{\"o}p-Tsutsui point interaction, high energy limit, $k \to \infty$
does not bring the system to classical limit. 
As is well-known, high energy limit of $\delta$ and $\delta^\prime$ interactions
corresponds to the free pass and perfect bouncing wall respectively,
thus supplying two legitimate
deterministic classical limits of a point interaction.
With scale invariant point interactions, we are presented with semi-transparent wall 
with finite penetration probability at all energies. 
Therefore, if we were to identify the high energy limit as a classical limit,
we are forced to consider stochastic dynamics whose randomness originates
directly from the probabilistic nature of quantum mechanics itself.

Irrespective to the problem of classical limit and correspondence,
our analytical expressions, made possible by the scale invariance of 
F{\"u}l{\"o}p-Tsutsui point interaction, 
shed light on how irregular quantum dynamics
emerge as the infinite-period limit of multi-periodic scattering matrices,
just as chaotic classical dynamics emerge as the infinite-period limit of 
multi-periodic motion.
In this connection, it should be useful to consider a complementary approach 
of trace-formula based analysis to our model.
With appropriate modifications, existing semiclassical treatments 
of quantum graphs \cite{DJ02} appear capable of handling the problem,
and the comparison to the current approach should yield further insight
into the singular and irregular dynamics in quantum mechanics. 

Lastly, we emphasize the utility of scale invariant interaction 
in the studies of aspects of
one-dimensional systems other than the current example of stochastic dynamics,
such as the spectral properties of regular lattice, in which analytical  expressions 
(\ref{SCECD})-(\ref{BOE}) are expected to become very convenient.
\\

We acknowledge our gratitude to Dr. I. Tsutsui, Dr. T. F{\"u}l{\"o}p, Dr. P. Seba, 
Dr. K. Takayanagi and Dr. T. Yukawa
for the enlightening discussions.
One of the authors (TC) thanks the members of the Theory Group at 
High Energy Accelerator Research Organization (KEK) for their
hospitality during his sabbatical stay.


%

\end{document}